\documentclass[prl,twocolumn,superscriptaddress,aps,showpacs]{revtex4}

\usepackage{here}
\usepackage{graphicx}
\usepackage{bm}

\newcommand{\be}{\begin{equation}}
\newcommand{\ee}{\end{equation}}

\newcommand\pictc[5]{\begin{figure}[t]
                       \centerline{\vspace{-2mm}
                       \includegraphics[width=0.9\columnwidth]{#3}}
                       \protect\caption{\protect\label{fig:#4} #5}\vspace{-2mm}
                    \end{figure}            }
\newcommand\pict[4][1]{\pictc{#1}{!tb}{#2}{#3}{#4}}
\newcommand\rpict[1]{\ref{fig:#1}}

\newcommand\leqt[1]{\protect\label{eq:#1}}
\newcommand\reqtn[1]{\ref{eq:#1}}
\newcommand\reqt[1]{(\reqtn{#1})}

\newcounter{Fig}

\begin{document}

\title{Reduced-symmetry two-dimensional solitons in photonic lattices}

\author{Robert Fischer}
\affiliation{Nonlinear Physics Centre and Laser Physics Centre, Centre for Ultrahigh-bandwidth Devices for Optical Systems (CUDOS), Research School of Physical Sciences and Engineering, Australian National University, Canberra, ACT 0200, Australia}

\author{Denis Tr\"ager}
\affiliation{Nonlinear Physics Centre and Laser Physics Centre, Centre for Ultrahigh-bandwidth Devices for Optical Systems (CUDOS), Research School of Physical Sciences and Engineering, Australian National University, Canberra, ACT 0200, Australia}
\affiliation{Institut f\"ur Angewandte Physik, Westf\"alische Wilhelms-Universit\"at, Corrensstr. 2, 48149 M\"unster, Germany}

\author{Dragomir N. Neshev}
\author{Andrey A. Sukhorukov}
\author{Wieslaw Krolikowski}
\affiliation{Nonlinear Physics Centre and Laser Physics Centre, Centre for Ultrahigh-bandwidth Devices for Optical Systems (CUDOS), Research School of Physical Sciences and Engineering, Australian National University, Canberra, ACT 0200, Australia}

\author{Cornelia Denz}
\affiliation{Institut f\"ur Angewandte Physik, Westf\"alische Wilhelms-Universit\"at, Corrensstr. 2, 48149 M\"unster, Germany}

\author{Yuri S. Kivshar} 
\affiliation{Nonlinear Physics Centre and Laser Physics Centre, Centre for Ultrahigh-bandwidth Devices for Optical Systems (CUDOS), Research School of Physical Sciences and Engineering, Australian National University, Canberra, ACT 0200, Australia}

\pacs{42.65.Jx, 42.65.Tg, 42.65.Sf}

\begin{abstract}
We demonstrate theoretically and experimentally a novel type of localized beams supported by the combined effects of total internal and Bragg reflection in nonlinear two-dimensional square periodic structures. Such localized states exhibit strong anisotropy in their mobility properties, being highly mobile in one direction and trapped in the other, making them promising candidates for optical routing in nonlinear lattices. 
\end{abstract}

\maketitle

\noindent

Nonlinear periodic structures attract increasing attention due to the possibility to engineer the transmission and reflection properties of waves, opening new applications of photonic crystals for all-optical signal processing and switching~\cite{Slusher:2003:NonlinearPhotonic}. An important part of such applications is the ability to control wave propagation in the form of nonlinear localized modes or solitons~\cite{Kivshar:2003:OpticalSolitons}.
Conventional approaches are based on wave transport through structural defects or embedded waveguides, however increased flexibility can be achieved when the internal structure and symmetry of the nonlinear state itself select the direction of propagation in defect-free periodic structures. 

The symmetry of solitons is intrinsically defined by the physical mechanisms responsible for light localization, and two distinct scenarios are possible in periodic photonic structures with self-focusing nonlinearity. First, it is light waveguiding in the regions of high refractive index due to the effect of total internal reflection; this effect leads to the formation of {\em discrete lattice solitons}~\cite{Christodoulides:2003-817:NAT, Eisenberg:1998-3383:PRL, Fleischer:2003-23902:PRL, Fleischer:2003-147:NAT, Neshev:2003-710:OL, Martin:2004-123902:PRL}. Second, it is resonant Bragg or Laue reflection in periodic structures; this effect leads to the formation of {\em gap solitons} with the frequencies inside the photonic bandgaps of one-dimensional~\cite{Slusher:2003:NonlinearPhotonic, deSterke:1994-203:ProgressOptics, Sukhorukov:2003-31:IQE, Fleischer:2003-23902:PRL, Mandelik:2004-93904:PRL, Neshev:2004-83905:PRL} and two-dimensional~\cite{John:1993-1168:PRL+, Mingaleev:2001-5474:PRL, Xie:2003-213904:PRL, Fleischer:2003-147:NAT} periodic structures. 

In this Letter, we demonstrate that in two-dimensional periodic nonlinear systems it is possible to utilize both localization mechanisms jointly to obtain self-trapped states with different properties along the two principal directions in a square lattice. We describe theoretically the families of such {\em reduced-symmetry gap solitons} and study experimentally the nonlinear beam self-trapping. Further, we identify the unique mobility of these gap solitons which is highly anisotropic along the two principal axes of the square lattice, suggesting a novel mechanism for directional nonlinear wave transport in symmetric lattices.

Optically-induced photonic lattices provide an important experimental tool to study the effects of nonlinearity on light propagation in periodic structures, being dynamically reconfigurable and possessing strong nonlinear response at the mW level. We study the propagation of an optical beam through a two-dimensional (2D) optically-induced lattice which is governed by the parabolic equation for the slowly varying amplitude of the electric field~\cite{Efremidis:2002-46602:PRE, Fleischer:2003-147:NAT, Neshev:2003-710:OL}
\begin{equation} \leqt{nls}
   i \frac{\partial E}{\partial z}
   + D \left( \frac{\partial^2 E}{\partial x^2}
               +\frac{\partial^2 E}{\partial y^2} \right)
   + {\cal F}(x,y, I) E = 0,
\end{equation}
where $D$ is the diffraction coefficient and the function ${\cal F}$ describes the refractive-index modulation induced by the lattice and the beam. We consider a periodic potential imprinted in a biased photorefractive crystal by an intensity pattern of the form ${\cal F}( x, y, |E|^2) = - \gamma (I_b + I_p(x) + |E|^2)^{-1}$. Here $I_b=1$ is the normalized constant dark irradiance, $I_p(x)=I_g [\cos(\pi (x+y) / d)+\cos(\pi (x-y)/d)]^2$ is the interference pattern [see Fig.~\rpict{bandgap}(b)] which induces a lattice with the period $d$, $\gamma$ is the nonlinear coefficient proportional to
the applied DC field, and $I=|E|^2$ is the intensity of the probe beam.

The propagation of linear (small-amplitude) waves is described by the spatially extended eigenmodes in the form of Bloch waves~\cite{Russell:1995-585:ConfinedElectrons}. The Bloch waves are found as solutions of the linearized Eq.~\reqt{nls} in the form $E = \psi(x,y) \exp( i \beta z + i K_x x + i K_y y)$, where $\psi(x,y)$ has the periodicity of the underlying lattice. For a square lattice the dispersion dependencies $\beta( K_x, K_y )$ are periodic with respect to translations $K_{x,y} \rightarrow K_{x,y} \pm 2 \pi / d$, and therefore are fully defined by their values in the first Brillouin zone. Then the calculated bandgap spectrum can be folded along the principal axis of the lattice, and it is shown in Fig.~\rpict{bandgap}(a). The parameters used for the calculations are chosen to match the experimental conditions: the diffraction coefficient is $D = z_s \lambda / (4 \pi n_0 x_s^2)$; $x_s=y_s=1\mu$m and $z_s=1$~mm are the scales of the dimensionless variables $x$, $y$, $z$; $n_0 = 2.35$ is the refractive index of a bulk photorefractive crystal, $\lambda = 532$~nm is the laser wavelength in vacuum, $\gamma = 2.36$, and $I_g=0.49$.

\pict{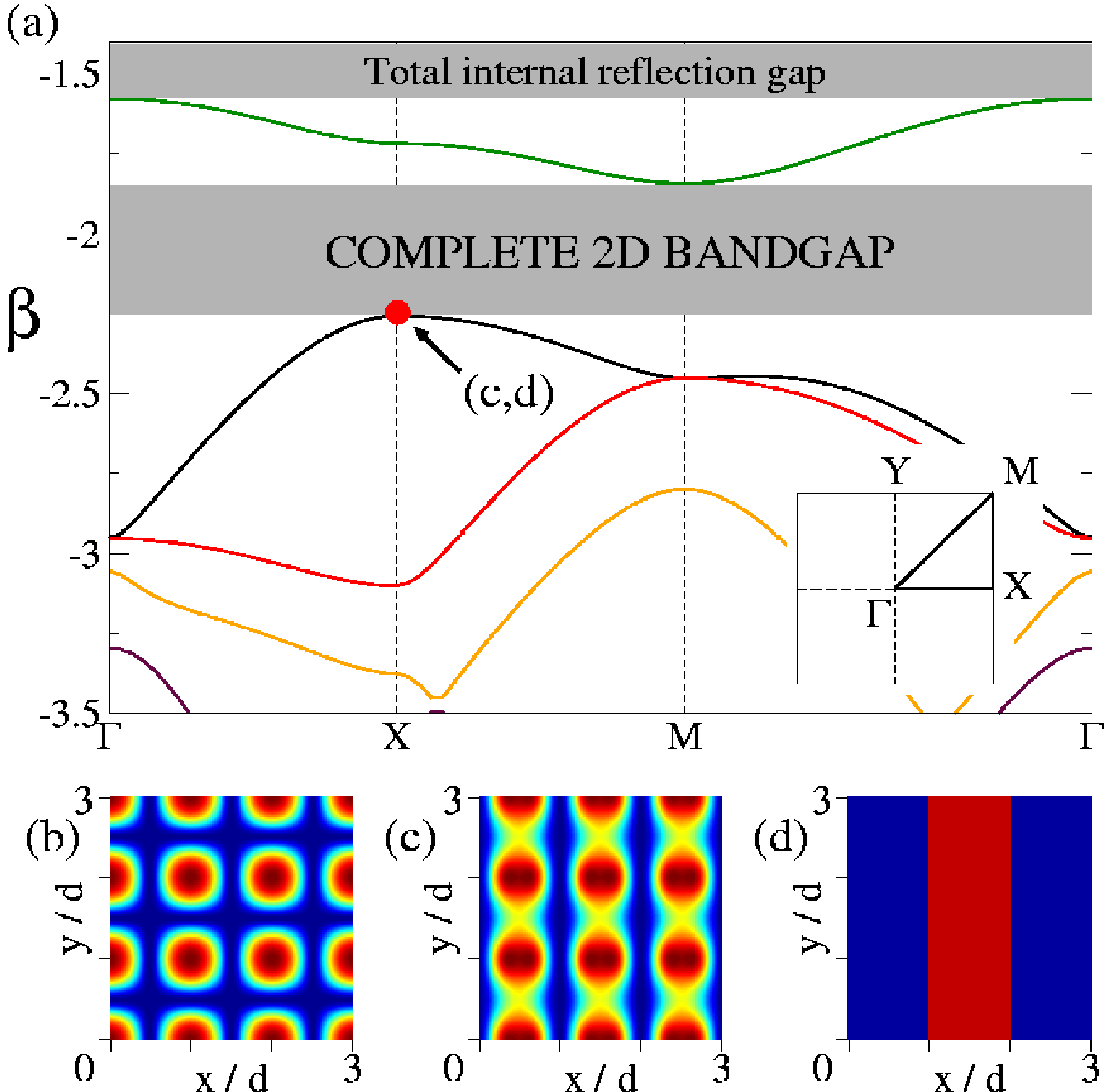}{bandgap}{(color online) (a,b) Bandgap spectrum and intensity distribution of the 2D square lattice. (c,d) Calculated intensity and phase distribution of the Bloch wave corresponding to the $X$-point of the second band of the spectrum (a).}

In a self-focusing medium nonlinearity increases the beam propagation constant, shifting it inside the gap for the modes corresponding to the top of dispersion bands, and allowing for the formation of self-trapped states or spatial solitons~\cite{Kivshar:2003:OpticalSolitons}. It follows that for a square lattice the solitons can originate from the $\Gamma$ point of the first band and $X$ (or $Y$) point of the second band~\cite{John:1993-1168:PRL+} [see Fig.~\rpict{bandgap}(a)]. At the $\Gamma$ point, the Bloch-wave diffraction is fully {\em symmetric} along the principal lattice directions ($x$ and $y$), giving rise to discrete lattice solitons~\cite{Fleischer:2003-147:NAT, Martin:2004-123902:PRL}. Symmetric superposition of $X$ and $Y$ states gives rise to symmetric gap solitons~\cite{John:1993-1168:PRL+, Mingaleev:2001-5474:PRL} or gap vortices~\cite{Bartal:2005-53904:PRL+}. It was predicted~\cite{John:1993-1168:PRL+}, that solitons may be also associated with pure $X$ (or $Y$) states, having different widths along the principal axis directions. The  analysis in Ref.~\cite{John:1993-1168:PRL+} was performed using the semi-analytical approach based on the coupled-mode theory, which can only provide qualitative description of wide solitons which extend over many lattice sites. However, we find that wide solitons may experience quasi-collapse and transform into highly-localized states. We demonstrate below that their properties strongly depend on the position with respect to the lattice, and show that this has a fundamental effect on the soliton transport.

\pict{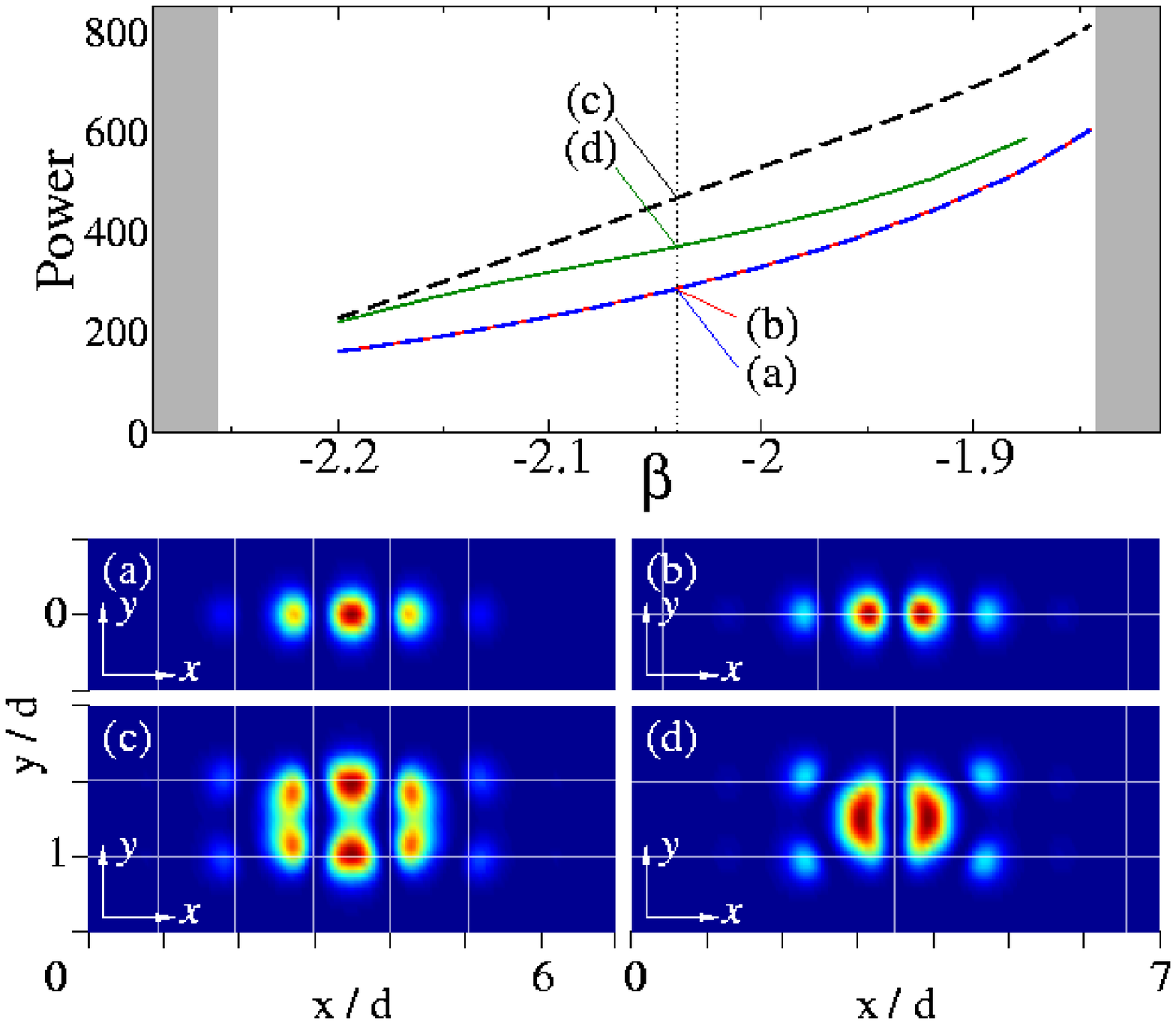}{families}{(color online) Families of gap solitons shown as normalized soliton power vs. the propagation constant. (a-d) Examples of four types of gap solitons for $\beta=-2.04$. Lines show the position of the lattice maxima.}

The families of gap solitons originating from the $X$-symmetry point are found numerically by the 2D relaxation technique, and they are shown in Fig.~\rpict{families}(top). There exist {\em four different symmetries} of the 2D gap solitons, as shown in the examples (a-d) for $\beta=-2.04$. They are centered at or between the lattice sites for the horizontal and vertical axis, respectively. In all the cases, the soliton phase is constant in $y$- and oscillating in the $x$-direction, resembling the structure of the corresponding Bloch state [see Fig.~\rpict{bandgap}(c,d)]. Alternating phase is a key signature for localization due to Bragg reflection, whereas the constant phase in the $y$-direction indicates that total internal reflection is responsible for self-trapping along this axis. We also note that the solitons are broader in the $x$- and more localized in the $y$-direction, having {\em reduced-symmetry} with respect to the underlying square lattice. This happens because the diffraction is stronger in the $x$-direction due to Bragg scattering, similar to the effect of dispersion enhancement in Bragg gratings~\cite{Slusher:2003:NonlinearPhotonic}, whereas along the $y$-direction, diffraction is weaker resembling discrete field tunneling in waveguide arrays~\cite{Christodoulides:2003-817:NAT}. 

The ability to control soliton motion in nonlinear lattices may open new possibilities for optical switching. It was shown that solitons can become trapped in a 1D photonic structure due to a self-induced Peierls-Nabarro potential when the optical power exceeds a certain threshold~\cite{Kivshar:1993-3077:PRE, Morandotti:1999-2726:PRL}. Most remarkably, we find that reduced-symmetry solitons can simultaneously be trapped in one direction, and fully mobile in another direction, offering a novel approach for control of nonlinear transport in symmetric lattices. 
Indeed, since the powers for the types (a) and (b) solitons depicted in Fig.~\rpict{families}(top) coincide for any value of the propagation constant, it follows that the Peierls-Nabarro potential is negligible along the $x$ axis, allowing for soliton propagation in this direction. On contrary, there is a large difference in the powers for the solutions of types (a,b) and (c,d). This implies the existence of a significant Peierls-Nabarro potential along the $y$ axis and
reduced mobility or trapping of the solitons along this direction. 
We also note that 
the states (a) and (b) appear energetically preferable and likely to be experimentally observable.

\pict{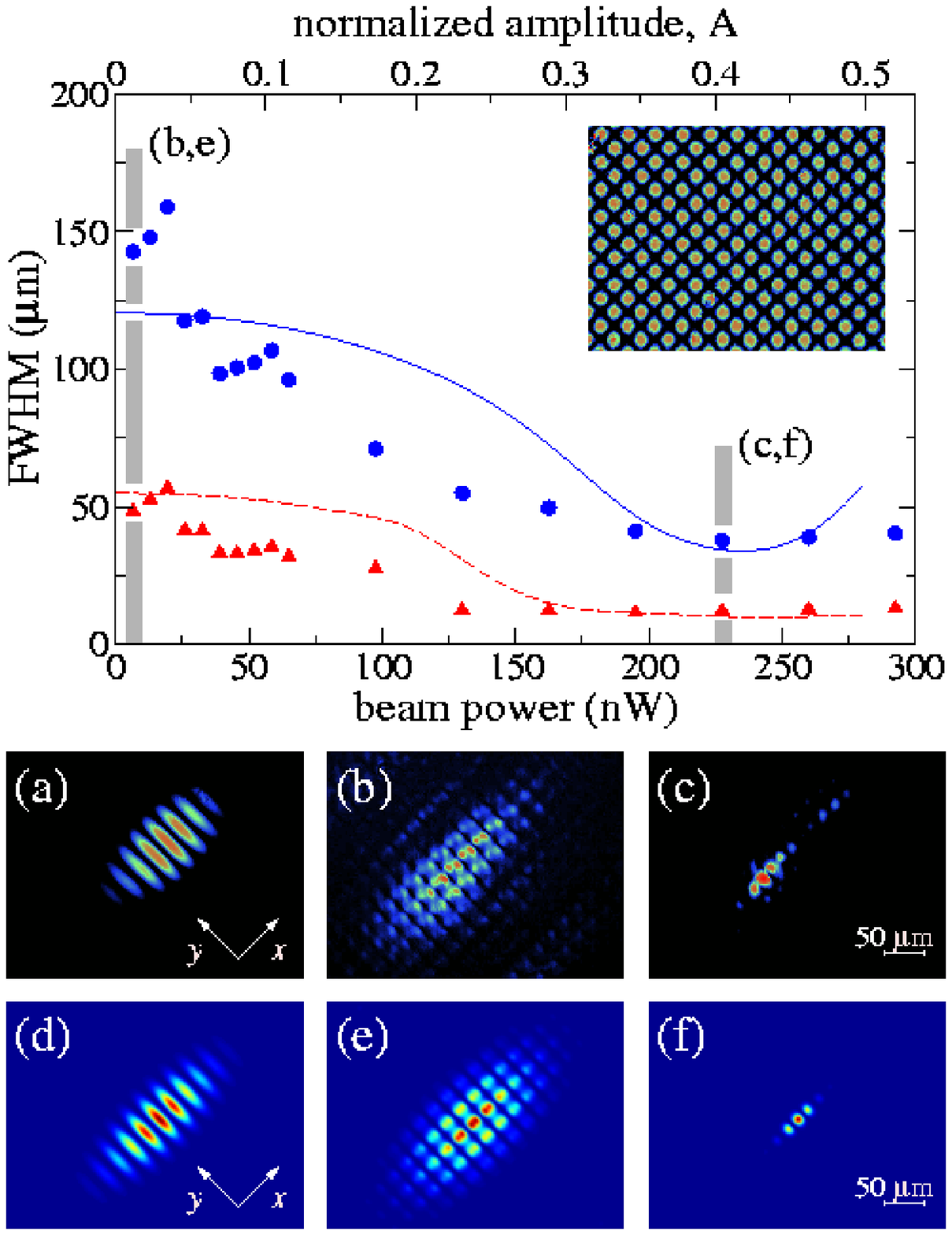}{widthplot}{(color online) Focusing of a beam corresponding to the {X}-point of the second band: (top) width at the crystal output in experiment (dots) and theory (lines). Solid line and dots -- widths along x; Dashed line and triangles -- widths along y. (a-c) experimental, (d-f) numerical results: (a,d) input beam profile; (b,e) output beam profiles at low power; (c,f) localized states at high power. Inset - optical lattice.}

In experiment, we induce a 2D periodic modulation of the refractive index in a biased 20~mm long SBN:60 photorefractive crystal~\cite{Efremidis:2002-46602:PRE}. As a light source we use a cw frequency-doubled Nd:YVO$_4$ laser at 532~nm. The laser beam is split into two parts by a polarizing beam splitter forming ordinary and extraordinary waves with respect to the crystalline axes. To create the photonic lattice, the ordinary polarized beam passes through a diffractive optical element (DOE), splitting it into four coherent broad beams. The DOE is imaged by an optical telescope on the front face of the crystal, thus creating a 2D square interference pattern with a period of 23~$\mu$m. The lattice extends through the whole length of the crystal and is oriented at 45 degrees with respect to the bias (horizontal) electric field [Fig.~\rpict{widthplot}(top, inset)], in order to reduce the effect of the intrinsic anisotropy of the photorefractive nonlinearity~\cite{Desyatnikov:2005-869:OL}. The second extraordinary-polarized beam is directed onto a programmable phase modulator (1024$\times$768 pixels), to allow for phase and amplitude engineering of the $X$-point Bloch states from the second band. The field structure produced by the modulator is also imaged onto the front face of the crystal and combined with the lattice onto a beam splitter. The front and the back faces of the crystal can be imaged onto a CCD-camera to capture beam intensity distribution.

In order to match the field profile of the linear Bloch state [Figs.~\rpict{bandgap}(c,d)], we use a broad input beam with several out-of-phase humps [Fig.~\rpict{widthplot}(a)], positioned in between the lattice sites along the $x$-direction, while centered onto a lattice site in the $y$-direction. To compensate for the anisotropic diffraction, the input beam is made elongated along the $x$-direction. Such a broad beam allows for {\em spectrally pure excitation} of the Bloch state from the top of the second band at the $X$ point, which can then be moved adiabatically into the gap as the beam self-focuses, thus avoiding radiation into other spectral bands.
In 2D lattices, the broad beams are often susceptible to collapse instability, as their width gradually decreases due to self-focusing when the input power slightly exceeds the threshold for the soliton formation~\cite{Mingaleev:2001-5474:PRL}. We find that for reduced-symmetry states, the collapse is arrested when the beam becomes confined to a single lattice site in the $y$ direction, effectively {\em reducing the dimensionality to a quasi-1D system} where collapse does not occur, similar to pulse dynamics in fiber arrays~\cite{Aceves:1995-73:PRL+}. 
The process of localization of the optical beam with increasing of the laser power is summarized in Fig.~\rpict{widthplot}. The predicted anisotropy in diffraction can be clearly seen in Fig.~\rpict{widthplot}(top), where the beam widths along the two lattice axes are plotted against the input power. The widths are estimated as the full width of the half maximum (FWHM) of the Gaussian envelope of the modulated beam profile in both directions. At low intensities [20~nW, Fig.~\rpict{widthplot}(b)], the beam diffracts linearly and its width increases more in $x$- than in $y$-direction despite the fact that the input size along the $x$-direction is larger. The spatial structure of the output beam well represents the characteristic fine-structure of the Bloch wave from the $X$-point of the second band, with out-of-phase lobes along the $x$ direction as confirmed with interferometric measurements. At higher laser powers, the beam focuses until a strongly localized state is formed at about 230~nW [Fig.~\rpict{widthplot}(c)] which covers approximately three lattice sites in the $x$ and one in $y$ directions.

To support our experimental results and confirm the formation of localized gap states we performed direct numerical simulations of Eq.~\reqt{nls} with the initial conditions similar to the experimental ones [cf. Figs.~\rpict{widthplot}(a) and~(d)], defining the input beam envelope as
$\psi(x,y) = A \exp(-x^2/\delta_x^2-y^2/\delta_y^2) \sin(2\pi x/d)$, 
where $\delta_x$ and $\delta_y$ are the beam widths along the two lattice
axes. We find that the results of the corresponding numerical simulations [lines in Fig.~\rpict{widthplot}(top) and Figs.~\rpict{widthplot}(e,f)] show the same tendencies as our experiments. Similar to the experiment, the width of the output state decreases to a minimum value of about 40~$\mu$m in the $x$-direction and down to one lattice period (23~$\mu$m) in the $y$-direction. The slight increase of the beam width along the $x$-direction at powers higher than 250~nW is due to beam splitting into two solitons. This effect is typical for one-dimensional solitons~\cite{Kivshar:2003:OpticalSolitons}, and shows that strongly localized reduced-symmetry solitons indeed demonstrate a quasi-1D dynamics. In the numerical simulations, we can also resolve the complete beam evolution inside the crystal. The calculations show that the beam experiences quasi-collapse at its initial stage of propagation, whereas it reaches a steady-state propagation afterwards. We find that the solitons may disintegrate due to the development of weak oscillatory instabilities at large propagation distances. However, all solutions from the family presented in Fig.~\rpict{families} propagate steadily for experimentally feasible distances. 

\pict{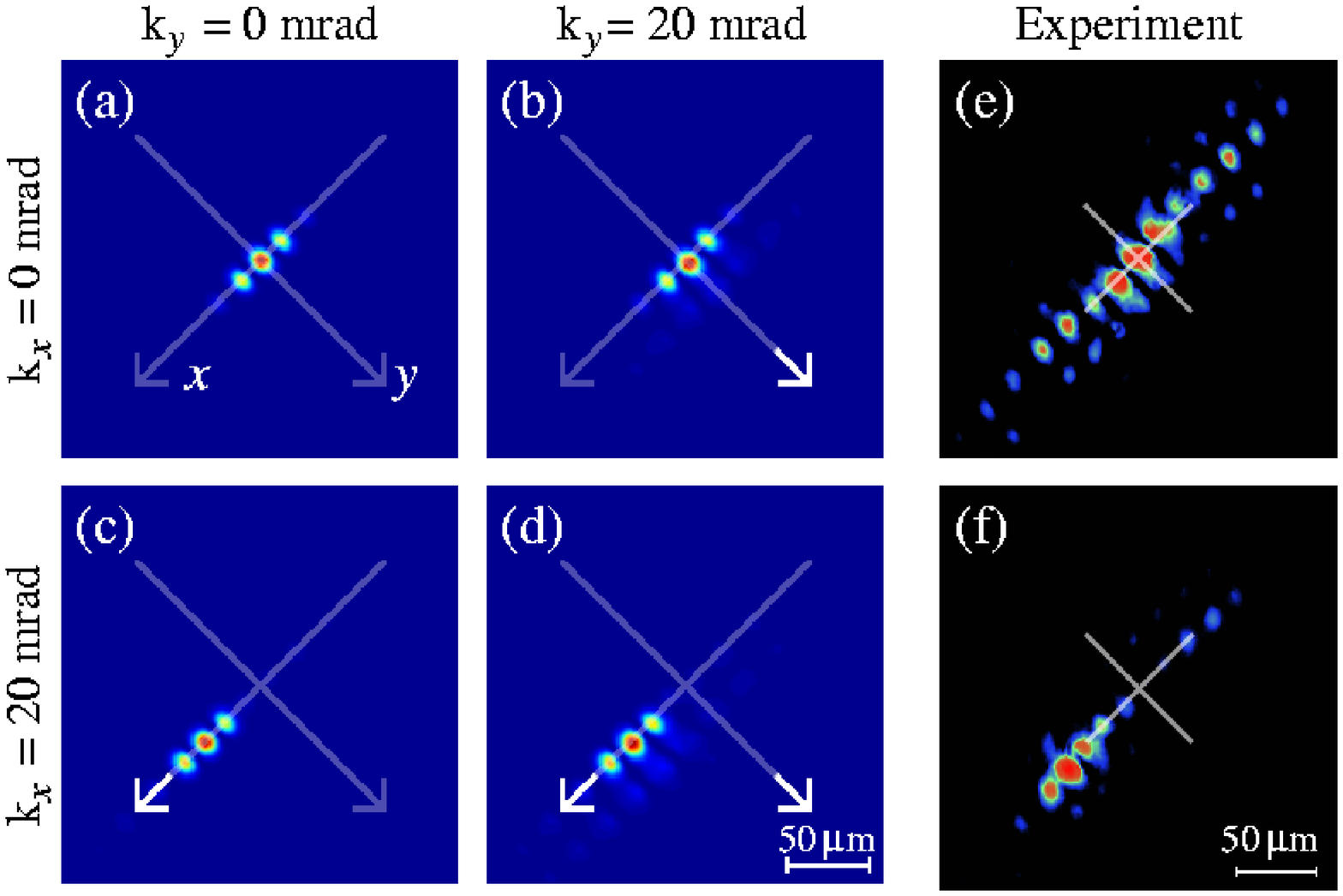}{mobility}{(color online) (a-d) Numerically calculated output beam profiles for different initial tilts of the beam. (a) no tilt, (b) 20~mrad tilt along $x$, (c) 20~mrad tilt along $y$, (d) 20~mrad tilt along $x$ and $y$. White bars mark lattice periods. (e,f) show two experimental examples for the mobility along $x$, the cross marking the beam center at the crystal front face.}

We now investigate the soliton mobility, which is expected to be highly anisotropic. We impose an initial tilt of the input beam along the different directions and study the beam displacement at the output, comparing to the immobile soliton position [Fig.~\rpict{mobility}(a)]. In simulations, an initial tilt of 20~mrad (15\% of the Bragg angle) in $x$ moves the output for two lattice sites [Fig~\rpict{mobility}(b)], whereas the same tilt in $y$ leads only to a small deformation but no movement of the output state [Fig~\rpict{mobility}(c)]. Even with both tilts superimposed, the soliton moves the same two lattice sites along the $x$ axis only [Fig~\rpict{mobility}(d)]. Two examples of the corresponding experimental results presented in Fig.~\rpict{mobility}(e,f) demonstrate the high mobility of the localized states along $x$ direction. We underline that the lattice itself is uniform in $x$ and $y$, and the directional mobility is only determined by the localized state itself. Their ability to robustly move along one row (or column for $Y$ state) of the lattice makes them good candidates for flexible soliton networks in 2D periodic structures. This offers new opportunities compared to the soliton networks suggested earlier for optical signal routing and switching~\cite{Christodoulides:2001-233901:PRL}.

In conclusion, we have presented the experimental observation of novel self-trapped gap states in a square two-dimensional lattice, where localization is based on the combined effects of total internal and Bragg reflection. These states originate from the $X$-symmetry point of the lattice bandgap spectrum, and they possess a reduced symmetry and highly anisotropic diffraction properties. Due to this anisotropy, they exhibit enhanced mobility along the modulated direction and are trapped by the lattice in the transverse direction, allowing for control of directional wave transport with possible applications for optical routing and switching in planar nonlinear periodic photonic structures.

This work has been supported by the Australian Research Council. D.T. thanks Nonlinear Physics Centre for hospitality during his stay in Canberra and Konrad-Adenauer-Stiftung e.V. for financial support.

\end{document}